\documentclass[twocolumn,prl]{revtex4}

\usepackage{epsf}
\usepackage{times}
\usepackage{graphicx}
\usepackage{bm}%

\begin{document}

\title{
Do All Spherical Viruses Have Icosahedral Symmetry?
}

\author{Eric Lewin Altschuler}
 \affiliation{Department of Physical Medicine \& Rehabilitation, UMDNJ\\
    University Hospital, 150 Bergen Street, B-403, Newark, NJ 07103, USA}
\email{eric.altschuler@umdnj.edu}    
    
\author{Antonio P\'erez--Garrido}
 \affiliation{ Departamento de F\'\i sica Aplicada, UPCT\\
Campus Muralla del Mar, Cartagena, 30202 Murcia, Spain}
\email{Antonio.Perez@upct.es}

\begin{abstract}
Recent high resolution structures for viral capsids with 12, 32  and 72 
subunits (${T1}$, ${T3}$ and ${T7}$ viruses) have confirmed theoretical predictions
of an icosadeltahedral structure with 12 subunits having five nearest neighbors (pentamers) and 
$(10T+2)-12$ subunits having six nearest neighbor subunits (hexamers).  Here we note that theoretical considerations 
of energy strain 
for ${T4}$,  ${T9}$, ${T16}$ and ${T25}$ viruses by aligned pentamers and energy strain along with the sheer number of possible arrangement of pentamers as the number of subunits grows, and simulations for such numbers of subunits make an icosadeltahedral 
configuration either miraculously unlikely or indicate that there must be a principle of capsid assembly of unprecedented fidelity in Nature.  
We predict, for example, that high resolution data will show $T4$ capsids to have $D_{5h}$
not icosahedral symmetry.
\end{abstract}

\maketitle

 More than half a century ago Crick and Watson \cite{CW56}  had the ingenious insight that viral capsids must be made of 
 multiple units of the same small number of proteins, lest the viral genome be orders of magnitude too large---if coding for 
 each of the hundreds or thousands of capsid proteins separately---to fit inside the capsid.  Caspar and Klug\cite{CK62} made 
 a significant  advance in appreciating that the structure of a number of viral capsids had icosahedral symmetry.  
 They described capsids 
 by a number $T= a^2 +b^2 +ab$ ($a$, $b$ non-negative integers)  having $N = 10T + 2$ subunits  arranged
 into an icosadeltahedral lattice. However, determination of the structure of these large capsids is a tour de force of 
 experimentation, and until 
 recent high resolution studies confirming
icosadeltahedral configurations for $T1$\cite{CP09} , $T3$\cite{CP09} and $T7$\cite{JC06,GG09}  viruses,
 the structures were  
 typically determined by fitting relatively low
resolution experimental data to a model of a capsid with icosahedral symmetry (see ref.\ \cite{SS02} and refs. therein). 
 There has been important work using theory, modeling and simulations to try to understand how icosadeltahedral capsids
 form\cite{BG03,ZR04,KT09}.
Here using J.J. Thomson's problem \cite{Th04} of the arrangement of unit point charges on a
 sphere we quantify the thermodynamic and kinetic properties of icosahedral configurations. 
 We find that for $T1$, $T3$ and $T7$
viruses icosadeltahedral capsids are favored energetically and
occur the vast majority of times in simulations starting from
random configurations. Conversely, 
intriguingly, for $N = 42$, 92, 162 and 252 ($T4$, $T9$, $T16$, $T25$) 
the icosahedral configuration is not only neither the minimum energy nor the most
commonly found in our simulations starting random configurations, and 
never is found in our simulations at all. Indeed, for $N = 42$ ($T4$) the minimum
energy configuration has $D_{5h}$ symmetry and occurs in 98\% of runs. The icosadeltahedral configuration 
has a higher energy and is not found. 
We predict that $T4$, $T9$, $T16$ and $T25$ virus capsids do not have icosahedral symmetry.
If they are found in high resoultion experimental data, then
novel ideas are needed to explain such high fidelity assembly of
thermodynamically and  statistically unfavored configurations.
 
Over one hundred years ago J. J.Thomson \cite{Th04}  asked the question of the minimum
energy configuration of N unit point charges on (the surface of) a unit conducting
\cite{ EH91,Ed92,Ed93,AW94,EH95,EH97,AW97,PD97,Do96,DM97,PD97b,PM99,BN00,BC02,BB03,AP05,BN06,AP06,WU06}
sphere. Much theoretical, numerical and experimental work since then
has made considerable progress on Thomson's problem yielding interesting and nonobvious
results: 
For $N = 4$ charges the global minimum energy configuration is the geometrically symmetric configuration of a tetrahedron.
However, for $N = 8$ the minimum energy configuration is not
a cube, but rather an anticube---four charges arranged in a square parallel to the equatorial
plane in both the Northern and Southern hemispheres, but with the squares rotated
by 45 degrees with respect to each other. This configuration has a lower energy than a
cube as the rotation of the squares lowers the energy between nearest neighbor charges
between the two squares. The case of $N = 8$ also illustrates a general phenomenon
in Thomson's problem whereby the most symmetric configuration is not necessarily
the configuration of minimum energy. However, symmetry considerations can also be
a useful guide to finding minimum energy configurations. For  $12\le N\le 100$ (and
likely for the most part up to $N = 200$) attack of Thomson's problem by multiple
numerical and theoretical approaches and methods has likely found the minimum energy
configurations. In most cases there are exactly twelve charges with five nearest
neighbors---pentamers and the rest of the charges with six nearest neighbors---hexamers. Euler's theorem for convex polyogons---the number of vertices plus faces equals the number of edges plus 2 $(V + F = E + 2)$---has the result for points on a sphere that there must be at least twelve pentamers with the rest of the charges being hexamers or pentamer/septamer
pairs. Though, only for $N$ of the form 10($a^2 + b^2 + ab)$ + 2 (where $a\geq b\geq 0$) is
it possible for the twelve pentamers and the entire configuration to have icosahedral
symmetry. While for $N$ = 12, 32, 72, 122, 132, 192, 212, 272 and 282 (Table 1, Figure 1) the
icosahedrally symmetric configuration (an icosadeltahedral configuration) is the best
known energy minimum (and the presumed global energy minimum configuration),
for $N$ = 42, 92, 162, 252 (and then $T$ numbers larger than 282) the icosadeltahedral
configuration is not the global energy minimum. Instead configurations with exactly
twelve pentamers, but with the pentamers arranged in $D_{5h}$, $D_2$, $D_3$ and $C_2$ symmetries are the global energy minima for $N = 42, 92, 162$ and 252 respectively (Table I, Figure 2). 
These symmetries of global energy minima hold not only for the $1/r$ Coulomb potential, but for other representative electrostatic potentials as well (Table II).

 The reason that the icosadeltahedral configurations are not the global minima for
$N =$ 42, 92 and 162 ((2, 0), (3, 0) and 4, 0)) configurations is due to the energy cost of
the vertices of the pentamers being nearer to each other than in non-icosahedral configurations.
In general it seems that in seeking global energy minima for $N<$200 Nature
uses the general strategies of moving and rotating pentamers. In some cases either because
of pure geometrical constraints  \cite{Ed92} or just to minimize the energy, occasionally
a pentamer/ heptamer defect pair (dislocation defect in the language of elasticity) is
needed to achieve a global energy minimum.
As $N$ grows larger--- $\gtrsim 500$ or so---important papers by Dodgson and Moore \cite{Do96,DM97}
showed that the energy strain of the pentamers which is necessitated by the topology
of a sphere but which distort the pure hexagonal lattice that would be the energy minimum
on a flat sheet, is such that to lower the energy pentamer/heptamer defect pairs
are needed between all of the pentamers.
Now, as $N$ grows the number of local energy minima, $M(N)$ ($12\le N\le 112$)
was found to grow exponentially  \cite{EH95}:
\begin{equation}
M(N)\approx 0.382\times \exp \left( 0.0497N\right).
\end{equation}
Thus, if Nature is using an energy minimization strategy to find the configuration of
alignment of molecules in a viral capsid it would seem that as $N$ grows it will become
increasingly difficult if not impossible to find the global energy minimum configuration.
Furthermore, the number and relative depth and breadth of good local minima
could also be a constraint on kinetic strategies that Nature may use to find the ultimate
configuration.
In Table 1 we show the number of times we found the various local minima in runs
where we started  the charges from 5000 random configurations and then used standard
conjugate gradient methods to go to a local minimum. We see that for $N = 12,$ 32 and 72 the
minimum energy configuration and overwhelmingly the most common is in fact the icosadeltahedral
configuration. For $N = 42$, 92 and 162 however, the icosadeltahedral configuration is not only
not the minimum energy configuration, it is not reached from random configurations
ever.
In Figure 3 we show the number of local minima found in our simulations (solid
red circles) and those given by Eq. 1 versus $N$. For large $N$ the number of minima
found diverges from Eq. 1 since the number of initial runs is comparable to the number
of local minima. Similarly, we found that for a potential energy of $1/r^{0.5}$ and $1/r^3$
again the icosahedral configurations for $N = 42$, 92 and 162 are not global minima
and virtually never occur in the simulations (Table II).
For $N = $42 ($T4$) icosahedral and $D_{5h}$ symmetries look quite similar (Figures 2c
and 4). If we only take into account the balls' positions in Fig. 4, we can pass from the
left model to the right one just by rotating an hemisphere by an angle of 2$\pi$/10.

Based on these electrostatic model potentials we would predict that $T1$, $T3$ and
$T7$ virus would have icosadeltahedral configurations as recently found experimentally
\cite{CP09,JC06,GG09}. $T13$ viruses may have an icosadeltahedral configuration. We believe that
high resolution studies of $T4$ viruses will show a $D_{5h}$ configuration in particular and
not an icosadeltahedral configuration. We predict that $T9$, $T16$ and $T25$ viruses will
not be found to have an icosadeltahedral configuration though we do not have a clear
prediction for the structure of these viruses. Conversely, if a $T4$, $T9$, $T16$ or $T25$
virus were found to have an icosadeltahedral configuration given the essentially vanishing
possibility of this from energetic considerations or statistical considerations based
on electrostatic potentials, it would indicate a mechanical rule of assembly to be discovered
that is of expontentially good precision.  It is still also a mystery why nature seems to so prominently use capsids with $T$ numbers of subunits to the exclusion of other numbers of subunits.  
The energetics and statistics are so favorable for $N = 12, 32$ and 72 ($T$1, $T$3, $T$7) 
\cite{EH91,EH97,Ed92} that protein subunits consistent with these configurations must have emerged.  
The same factors would suggest a $D_{5h}$ configuration for $N$ = 42.  A key question could be understanding the evolution of $T$16 capsids. In general, geometry and topology seem to be important constraints that need to be considered in viral evolution or even possible treatments for viral diseases.

\newpage

\vglue 0.5cm

\begin{table*}
\caption{
Frequencies of finding global minima and icosadeltahedral configurations.
Erber and Hockney found similar frequencies using a diferent minimization algorithm\cite{EH91}}
\begin{tabular}{cccc}
\hline
N & Global minimum frequency  (E\&H freq.)& Icosadeltahedral frequency (E\&H freq.) & Symmetry of global minimum\\
\hline
12 & 100\% (100\%) & 100\% (100\%) & $I_h$\\
\hline
32 & 97.88\% (97.93\%) & 97.88\% (97.93\%)&$I_h$\\
\hline
42 & 98.18\% (98.08\%) &  0\% (0.03\%)& $D_{5h}$\\
\hline
72 & 83.84\%(82.95\%) &  83.84\% (82.95\%)&$I$\\
\hline
92 & 27.86\% (28.10\%)& 0\% (0\%)&$D_2$\\
\hline
132 & 23.82\%  & 23.82\%&$I$\\
\hline
162 & 0.84\% & 0\%& $D_3$\\
\hline
192 & 1.4\% & 1.4\%&$I$\\
\hline
212 & 0.26\%& 0.26\%&$I$ \\
\hline
252 & 0.08\%& 0\%& $C_2$\\
\hline
272 & 0\% & 0\%&$I_h$ \\
\hline
\end{tabular}
\end{table*}

\vglue 0.5cm

\begin{table*}
\caption{Frequencies of finding global minima and icosadeltahedral configurations.  Data for a $1/r^{0.5}$  
($1/r^{3}$) potential.}
\begin{tabular}{cccc}
\hline
N & Global minimum frequency  & Icosadeltahedral frequency& Symmetry of global minimum \\
\hline
12 & 100\% (100\%) & 100\% (100\%) &$I_h$\\
\hline
32 & 97.94\% (97.96\%)& 97.94\%  (97.96\%)&$I_h$\\
\hline
42 & 98.88\% (97.88\%) & 0.02\% (0.02\%) &$D_{5h}$\\
\hline
72 & 85.56\%  (81.44\%)&  85.56\% (81.44\%)& $I$\\
\hline
92 & 28.04\%  (27.96\%)& 0\% ( 0\%) & $D_2$\\
\hline
132 & 26.84\% (16.42\%)  &  26.84\% (16.42\%)& $I$\\
\hline
162 & 1.68\%  (0.4\%) & 0\% (0\%)& $D_3$\\
\hline
\end{tabular}
\end{table*}

\begin{figure}

\leavevmode
\includegraphics[angle=-90,width=8cm]{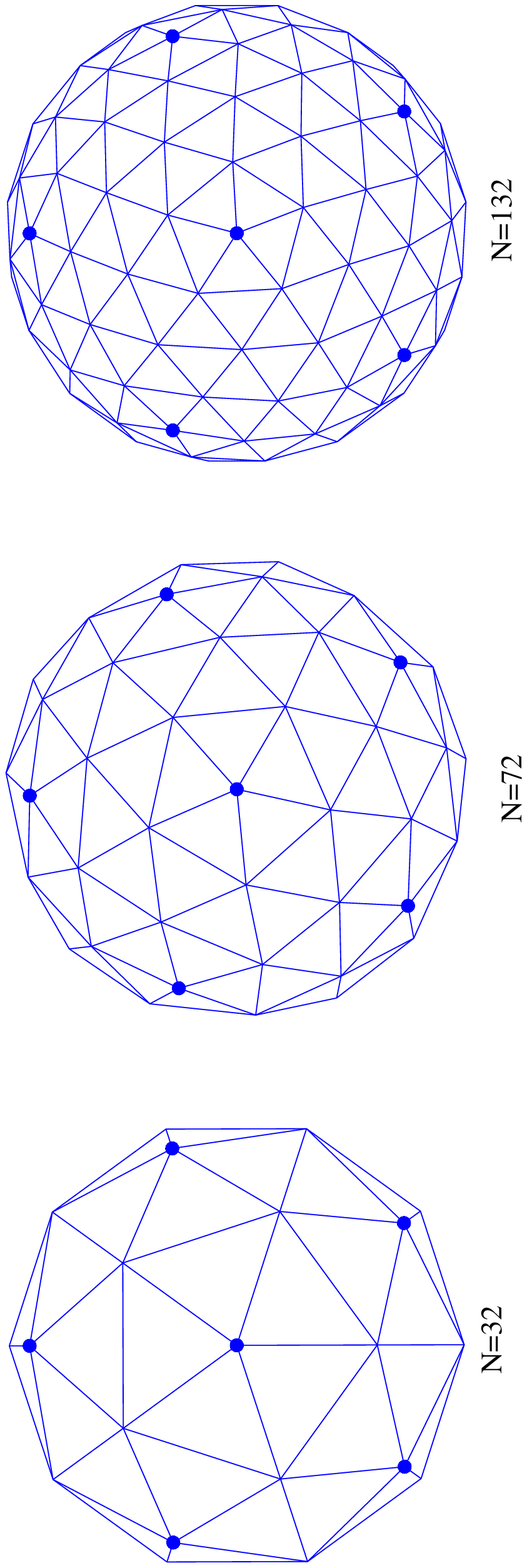}
\vglue -2cm
\caption.
{Icosadeltahedral global minima.}

\end{figure}

\begin{figure}

\leavevmode
\includegraphics[angle=-90,width=9cm]{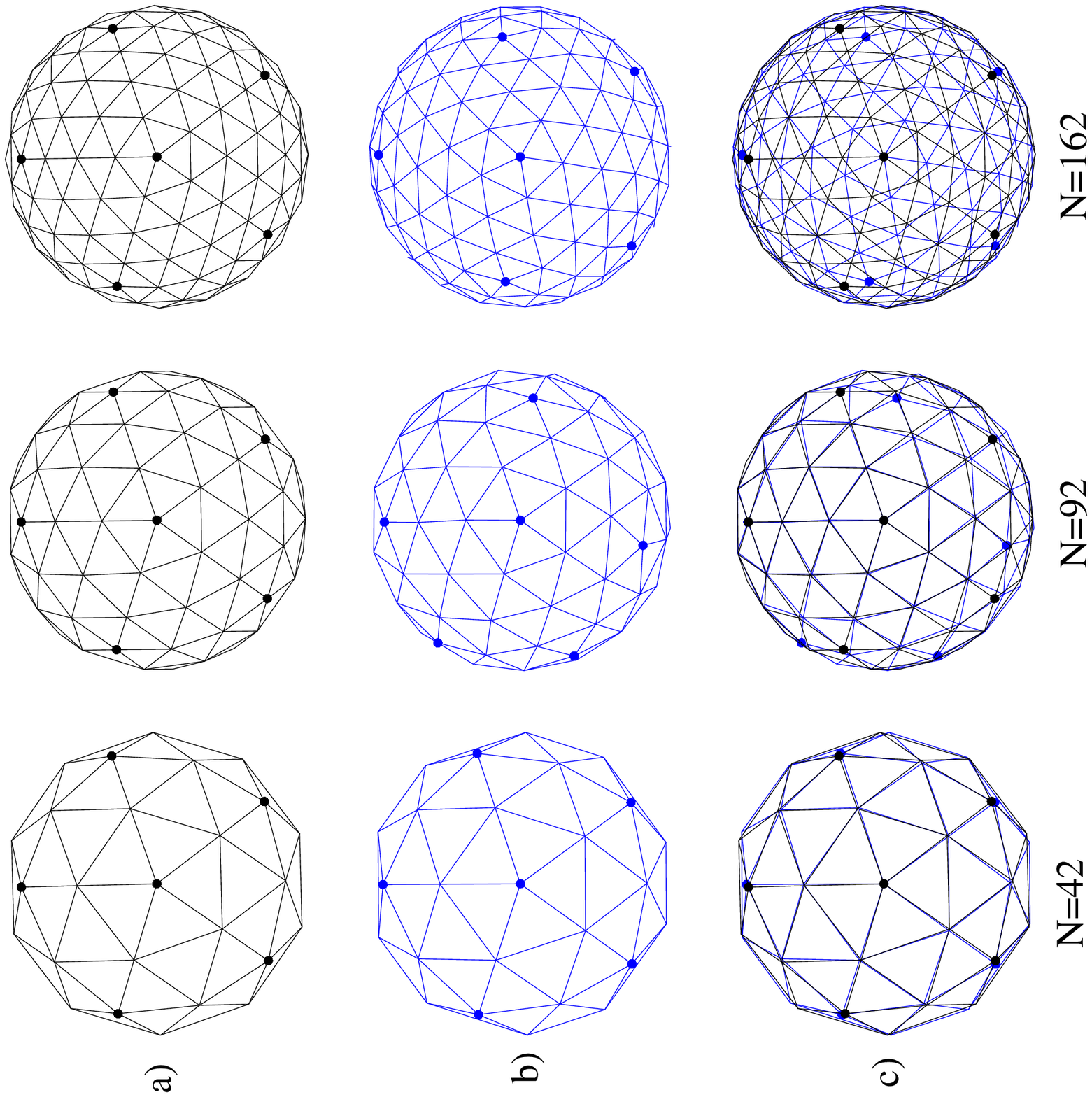}
\caption{
Icosadeltahedral configuration b) Global minima \hbox{c) Icosadeltahedral} and global minima overlap.}
\end{figure}

\begin{figure}

\leavevmode
\includegraphics[angle=-90,width=6cm]{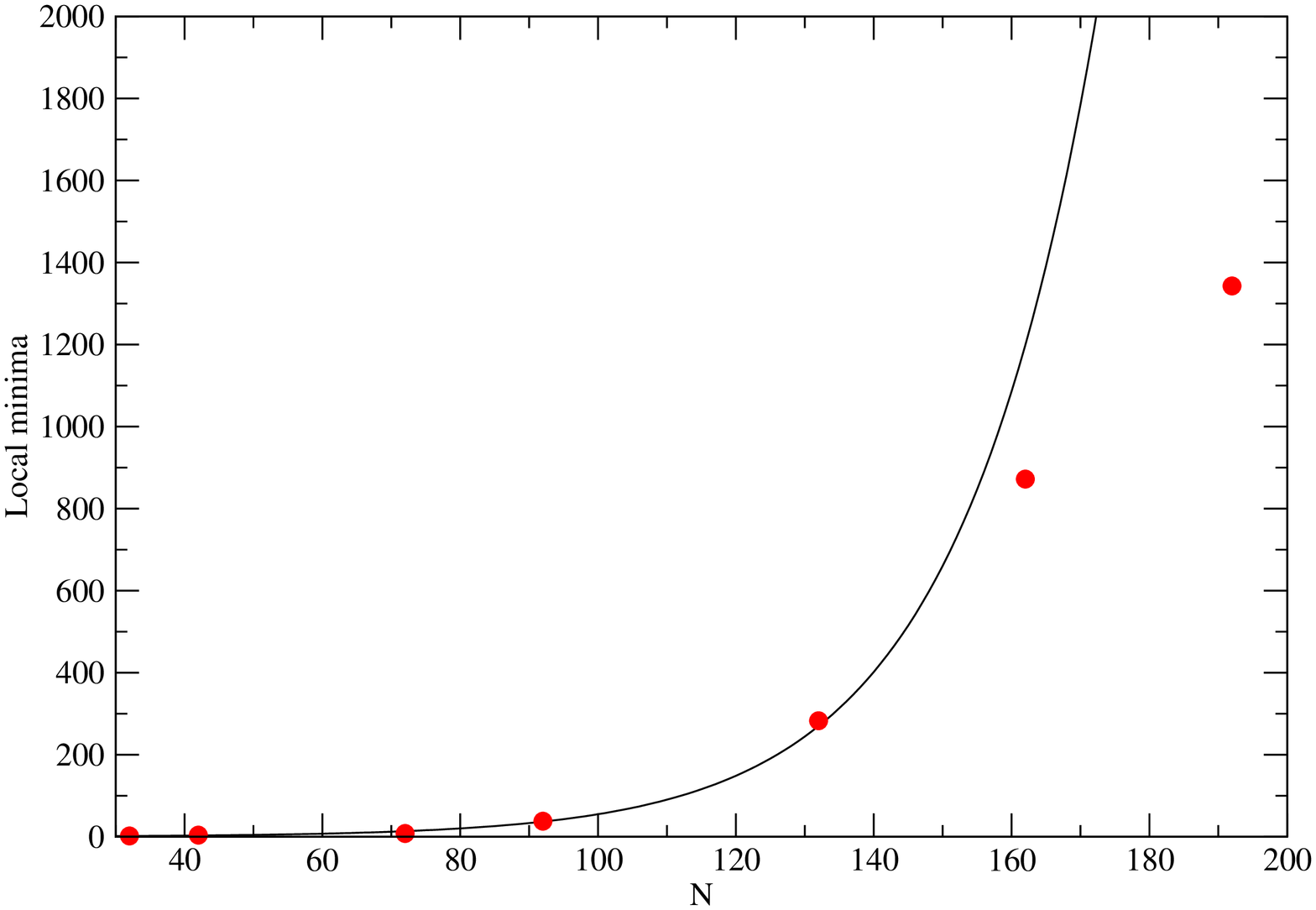}
\caption{
 Number of local minima found as a function of $\mathbf{N}$. The continuous line is given by Eq. 1.
} 
\end{figure}

\begin{figure}
\leavevmode
\includegraphics[angle=0,width=6cm]{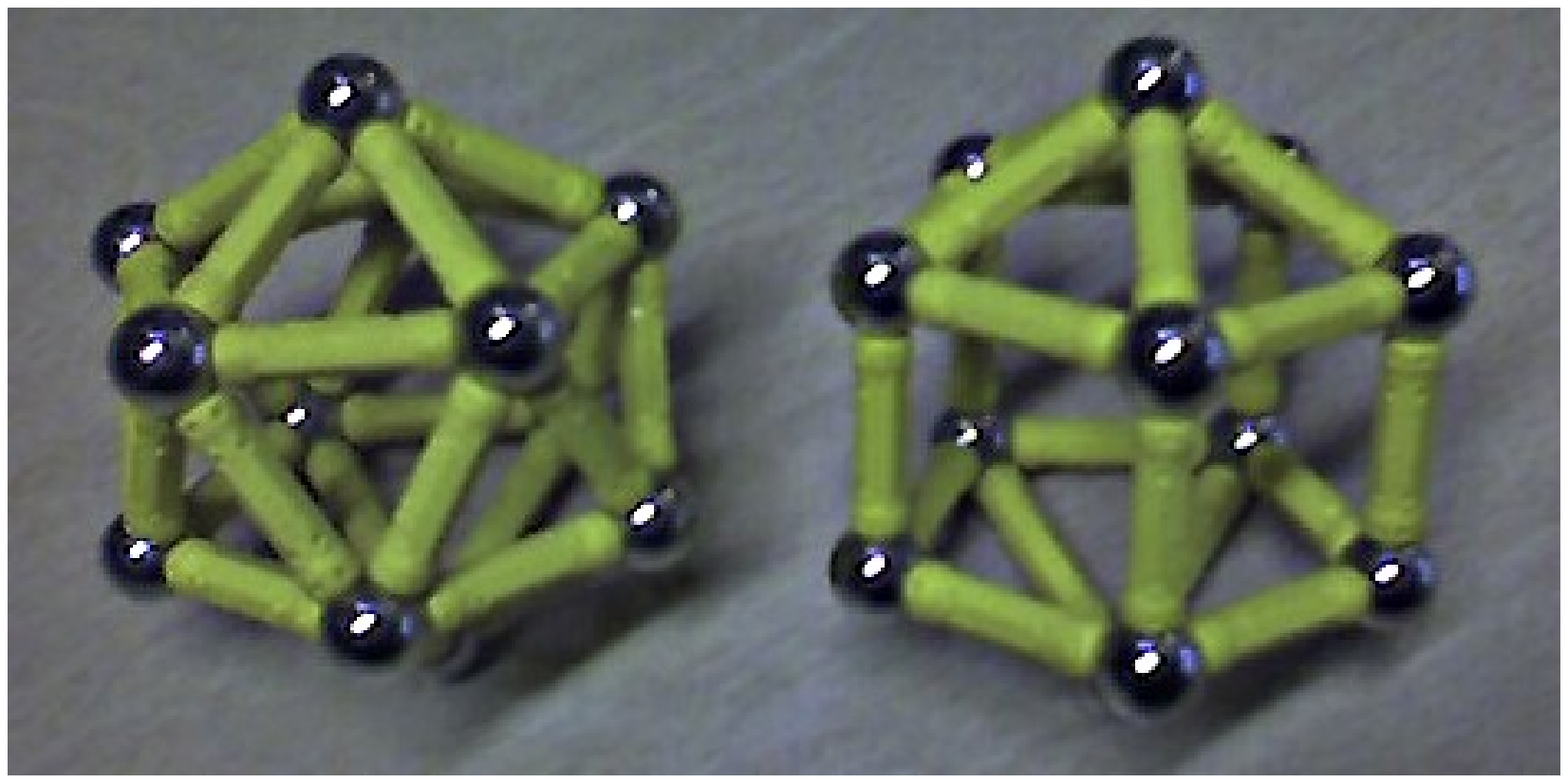}
\caption{
 Models of an icosahedron (left) and one with  $\mathbf{D_{5h}}$ symmetry (right) that resembles the global minimum for 
 $\mathbf{N=42}$.}
\end{figure}

 \end{document}